# Generalized Fock space and contextuality

Sergey Rashkovskiy[1,2], Andrei Khrennikov[3,4]

[1]*Ishlinsky Institute for Problems in Mechanics RAS, Vernadskogo Ave., 101/1, Moscow, 119526, Russia*

[2]*Tomsk State University, 36 Lenina Avenue, Tomsk, 634050, Russia*

[3]*International Center for Mathematical Modeling in Physics, Engineering, Economics, and Cognitive Science, Linnaeus University, 351 95 Växjö, Sweden*

[4]*National Research University for Information Technology, Mechanics and Optics (ITMO), department, St. Petersburg, 197101, Russia*

E-mail: Andrei.Khrennikov@lnu.se

**Abstract.** This paper is devoted to linear space representations of contextual probabilities - in generalized Fock space. This gives the possibility to use the calculus of creation and annihilation operators to express probabilistic dynamics in the Fock space (in particular, the wide class of classical kinetic equations). In this way we reproduce the Doi-Peliti formalism. The context-dependence of probabilities can be quantified with the aid of the generalized formula of total probability – by the magnitude of the interference term.

**Keywords:** Fock space; Doi-Peliti formalism; quantum probability; contextuality; interference of probabilities; formula of total probability; calculi of creation-annihilation operators, classical versus quantum dynamics, kinetic equation

1. ## Introduction

In this paper we construct linear space representations of contextual probabilities in generalized Fock spaces; in particular, our approach covers the quantum probability and Doi-Peliti probability formalism [1-5]. The present article is closely related to the series of works [6-12] on a quantum-like representation of contextual probabilistic models in complex Hilbert space. In such models, the degree of contextuality (for non-compound systems) was represented through perturbation of *the classical formula of total probability*:

$$P_A = p_1 p_A^{(1)} + p_2 p_A^{(2)},$$

by the magnitude of the *interference term:*

$$P_A = p_1 p_A^{(1)} + p_2 p_A^{(2)} + 2\sqrt{p_1 p_2 p_A^{(1)} p_A^{(2)}} \cos(\gamma_A),$$

where *A* is an event of observation of some fixed outcome and $p_A^{(i)}$=*P(A|i), i=1,2*, are contextual probabilities corresponding to observation-contexts labeled by *i*. The generalized Fock spaces introduced in the present paper can also serve for the same purpose. The linear space representations of contextuality can differ essentially from the standard quantum representation –

by the form of the interference term (see section 5). In particular, some representations generate the modifications of the formula of total probability with interference terms depending linearly on probabilities (see again section 5). Such representations are interesting as intermediate between classical (non-contextual) probability and quantum (contextual) probability. Our paper can have a quantum foundational impact by making smaller the gap between classical and quantum representations of probability.

We remind that interrelation of classical (measure-theoretic) and quantum (Hilbert space) probabilities is a very complex problem. Since the first days of creation of quantum mechanics experts in quantum foundations search the possibility to construct a classical probability representation of quantum processes. The common opinion that such a representation does not exist was originally based on the straightforward interpretation of the Heisenberg's uncertainty principle, negativity of the Wigner distribution, and later by various no-go theorems, e.g., the von Neumann, Kochen-Specker, and Bell theorems. However, the real situation is more complicated, see, e.g., Man'ko et al. [13-16], Khrennikov et al. [6-8, 17-19], and Dzhafarov et al. [20-25] See especially their recent works [13, 25, 19, 26]; see also papers on the attempts to reproduce quantum probabilities and correlations by using theory of classical random fields [27-35].

For special types of commutation relations for creation-annihilation operators, our linear space representation applied to the classical probabilistic dynamics induces realization of the wide class of *kinetic equations* in the Fock space. In particular, in this way we reproduce the Doi-Peliti formalism [1-4]. As shown in [1-4], such a formulation helps in understanding the relation between classical and quantum statistics, as well it provides new points of view on the nature of classical random processes. The Doi-Peliti formalism looks very similar to quantum mechanics (although it has nothing to do with it), and e.g. for continuous systems many of the concepts of quantum field theory become applicable. A very interesting application of this method might be *decision making theory*, cf. [36-45], see especially [44, 45] for applications of the calculus of creation-annihilation operators to decision making and game theory.

We remark once again that in this paper we concentrate on *contextual representations of non-compound systems*, cf. [17-19, 21, 22, 25, 26]. We also point out that the notion of contextuality used in the present paper differs from the notion which is typically discussed in literature on quantum foundations. The latter can be called "Bell-like contextuality''. It is adapted well to the scheme of the Bohm-Bell experiment and tests on violations of the Bell type inequalities.

We recall this notion. Let a, b, c be three quantum observables represented by Hermitian operators A, B,C. Let a is compatible with b and with c, i.e., [A, B]=0 and [A, C]=0. If the outputs of the a-measurements depend on whether it is measured jointly with b or with c, this phenomenon is called contextuality.

The contextuality notion explored in our paper goes back to the Copenhagen interpretation and Bohr's complementarity principle. By this principle all components of the experimental arrangement should be taken into account: outputs of measurements depend on the complete experimental context. It is impossible to split contributions of a system under measurement and the experimental context.[1] We can call this sort of contextuality *"Bohr-like contextuality"*. This notion is more general than Bell-like contextuality. The latter is reduced to the class of contexts determined by observables compatible with an observable a.

## 2. Rising-lowering (creation-annihilation) operators and Fock space

Consider the abstract Fock space, which is given by the basis $|n\rangle$, where $n$ are integers. The dual space to it is given by the basis $\langle n|$. By definition, the inner product of orts satisfies the condition

$$\langle n|k\rangle = \delta_{nk} \qquad (1)$$

Note that unlike [1-4] we use the natural definition of the inner product of the orts when the ort's norm is equal to one.

Arbitrary vector in this space

$$|u\rangle = \sum_{n=-\infty}^{\infty} c_n |n\rangle \qquad (2)$$

where $c_n$ are (in the general case) the complex parameters that are the "coordinates" of points in this space.

The vector which is dual to (2) is

$$\langle u| = \sum_{n=-\infty}^{\infty} b_n \langle n| \qquad (3)$$

where $b_n$ are (in the general case) the complex parameters that are the "coordinates" of points in the dual space.

Taking into account (1), for the norm of the vector (2) one obtains

$$\langle u|u\rangle = \sum_{n=-\infty}^{\infty} c_n b_n \qquad (4)$$

In the general case, the space $|n\rangle$ can be defined using the raising and lowering operators (creation and annihilation operators) acting on orta $|n\rangle$:

$$a^\dagger |n\rangle = f(n)|n+1\rangle \qquad (5)$$

---

[1] This irreducible dependence on the concrete experimental context motivates the assumption that not all contexts and hence not all quantum observables should be compatible – there can exist incompatible or complementary experimental contexts. The latter is the essence of the complementarity principle.

$$a|n\rangle = \varphi(n)|n-1\rangle \tag{6}$$

where $a^\dagger$ is the rising operator operator (creation operator); $a$ is the lowering operator (annihilation operator); $f(n)$ and $\varphi(n)$ are some functions.

Note that here $n$ are integers ($n \in \mathbb{Z}$), in contrast to the second quantization formalism and Doi-Peliti formalism [1-4], which, as will be shown later, are particular cases of the theory under consideration, and in which $n$ are only natural numbers ($n = 0,1,2,...$).

Taking into account (5) and (6), one finds the commutator

$$[a, a^\dagger] \equiv a\,a^\dagger - a^\dagger a = f(n)\varphi(n+1) - f(n-1)\varphi(n) \tag{7}$$

By specifying different functions $f(n)$ and $\varphi(n)$, one can obtain different Fock spaces. In the general case there are no restrictions on the functions $f(n)$ and $\varphi(n)$. Restrictions on these functions can be imposed by setting the form of the commutator (7).

Let be

$$[a, a^\dagger] = K(n) \tag{8}$$

where $K(n)$ is some function that determines the commutator.

Then, taking into account (7) and (8), one obtains the equation

$$f(n)\varphi(n+1) - f(n-1)\varphi(n) = K(n) \tag{9}$$

which determines the connection of the functions $f(n)$ and $\varphi(n)$ for a given function $K(n)$.

Denote

$$F(n) = f(n)\varphi(n+1) \tag{10}$$

Then we rewrite the equation (9) in the form

$$F(n) - F(n-1) = K(n) \tag{11}$$

From Eq. (11) we obtain the recurrence relation $F(n) = F(n-1) + K(n)$ from which it follows

$$F(n) = F_0 + \sum_{k=1}^{n} K(k)\,; n > 0 \tag{12}$$

$$F(n) = F_0 - \sum_{k=0}^{n+1} K(k)\,; n < 0 \tag{13}$$

where $F_0 = F(0)$. Thus, for a given value of $F_0$, the function $F(n)$ is uniquely determined by the function $K(n)$.

Then, from (10) one obtains

$$\varphi(n) = \frac{F(n-1)}{f(n-1)} \tag{14}$$

Thus, in this case the function $\varphi(n)$ is uniquely determined by the function $f(n)$.

The Fock space can be uniquely determined using the raising and lowering operators $a^\dagger$ and $a$, using the definitions (5) and (6):

$$|n\rangle = \begin{cases} \dfrac{a^{\dagger n}|0\rangle}{\prod_{k=0}^{n-1} f(k)}\,; n > 0 \\ \dfrac{a^{|n|}|0\rangle}{\prod_{k=0}^{n+1} \varphi(k)}\,; n < 0 \end{cases} \tag{15}$$

Expressions (5), (6), (12) - (15) completely determine the raising and lowering operators and the Fock space in the general case.

Consider a special case.

$$K(n) = 1 \tag{16}$$

i.e.

$$[a, a^\dagger] = 1 \tag{17}$$

In this case, taking into account (11) and (14), one obtains $F(n) = F_0 + n$ and

$$\varphi(n) = \frac{n + F_0 - 1}{f(n-1)}; (n \neq 0) \tag{18}$$

If we additionally take

$$F_0 = 1 \tag{19}$$

then we obtain

$$\varphi(n) = \frac{n}{f(n-1)} \tag{20}$$

In this case

$$a^\dagger a |n\rangle = n|n\rangle \tag{21}$$

That is, the operator

$$N = a^\dagger a \tag{22}$$

is an operator of integers: its eigenvalues are integers $n = \cdots, -2, -1, 0, 1, 2, \ldots$, while its eigenvectors are basis $|n\rangle$.

A special case (16), (19) corresponds to the definition of the creation and annihilation operators in the second quantization formalism and in the Doi-Peliti formalism [1-4]. To do this, we additionally assume that $n$ takes only non-negative values $0, 1, 2, \ldots$. Then the definition (6), which now takes place only for $n > 0$, must be supplemented by the condition $a|0\rangle = 0$.

In this case, assuming

$$f(n) = 1 \tag{23}$$

one obtains

$$\varphi(n) = n \tag{24}$$

$$|n\rangle = a^{\dagger^n}|0\rangle \tag{25}$$

which corresponds to Doi-Peliti formalism [1-4].

Assuming

$$f(n) = \sqrt{n+1} \tag{26}$$

one obtains

$$\varphi(n) = \sqrt{n} \tag{27}$$

$$|n\rangle = \frac{a^{\dagger^n}|0\rangle}{\sqrt{n!}} \tag{28}$$

which corresponds to the second quantization formalism for the bosonic field.

Thus, we see that it is possible to introduce a variety of other definitions of the Fock space, corresponding to the case of (16) and (19), which are different by the function $f(n)$. For all these cases, equation (21) which defines the basis $|n\rangle$, holds,

In order for the operators $a^\dagger$ and $a$ to be adjoint to each other, the condition $\langle k|a^\dagger|n\rangle = \overline{\langle n|a|k\rangle}$ which taking into account Eqs. (5) and (6) takes the form $f(n)\delta_{k,n+1} = \overline{\varphi(k)}\delta_{n,k-1}$.

It can be seen that this condition is satisfied only when $f(n) = \overline{\varphi(n+1)}$. This takes place, for example, in the case of second quantization (26), (27), while in the general case (including the Doi-Peliti formalism (23), (24)), the operators $a^\dagger$ and $a$ are not adjoint to each other. Thus, setting the functions $f(n)$ and $\varphi(n)$ allows defining a wider class of raising and lowering operators than that of using an adjoint condition.

### 3. Arbitrary operators in Fock space

Any operator $\widehat{W}$ acting on basis $|n\rangle$ can be expanded with respect to the basis $|n\rangle$.

In fact, the result of the action of the operator $\widehat{W}$ on the basis $|n\rangle$ can be expanded with respect to the basis $|n\rangle$:

$$\widehat{W}|n\rangle = \sum_k w_{nk}|k\rangle \tag{29}$$

where $w_{nk}$ are the decomposition coefficients.

Then, taking into account (1), the operator $\widehat{W}$ can be formally represented as

$$\widehat{W} = \sum_n \sum_k w_{nk}|k\rangle\langle n| \tag{30}$$

In particular, there is a linear operator $\hat{A}$ which acts on the basis $|n\rangle$, such that

$$\hat{A}|n\rangle = a_n|n\rangle \tag{31}$$

Thus, the value $a_n$ is an eigenvalue of the operator $\hat{A}$, while the vector $|n\rangle$ is the corresponding eigenfunction (eigenvector) of this operator.

Taking into account (30), the operator $\hat{A}$ can be formally represented as

$$\hat{A} = \sum_n a_n|n\rangle\langle n| \tag{32}$$

Taking into account (5) - (7), the raising and lowering operators can also be formally decomposed with respect to the basis $|n\rangle$:

$$a^\dagger = \sum_k f(k)|k+1\rangle\langle k| \tag{33}$$

$$a = \sum_k \varphi(k)|k-1\rangle\langle k| \tag{34}$$

An arbitrary operator $\hat{A}$ (31) can be expressed in terms of the operator $a^\dagger a$.

In fact, according to (5) and (6) one obtains $a^\dagger a|n\rangle = f(n-1)\varphi(n)|n\rangle$, or, taking into account (14), $a^\dagger a|n\rangle = F(n-1)|n\rangle$. This expression can be rewritten in the form $(a^\dagger a + C)|n\rangle = [F(n-1) + C]|n\rangle$, where $C$ is the arbitrary constant. Then, believing

$$F(n-1) = a_n - C \tag{35}$$

one obtains

$$\hat{A} = a^\dagger a + C \tag{36}$$

Representing the operator $\hat{A}$ in the form (36) we fix the form of the function $F(n)$ (35). Moreover, the function $f(n)$ can be arbitrary.

Note that in this way, one can express only one of the set of operators of type (31) defined in the space $|n\rangle$. All other operators of type (31) in the same space $|n\rangle$ with a fixed function $F(n)$ can also be expressed in terms of the raising and lowering operators, but this connection will be more complex than (36) and can be represented as a function $\hat{A} = \Lambda(a^\dagger, a)$, where $\Lambda(x, y)$ is some function that can be represented as a power series by its arguments.

We first consider a particular case: the representation of the operators $\hat{A}$ (32) through rising and lowering operators. We assume that the eigenvalues of the operator $\hat{A}$ can be written as a function of the integer $n$: $a_n = \alpha(n)$. In addition, we assume that the function $\alpha(n)$ can be expanded in a power series

$$\alpha(n) = \sum_{s=0}^{\infty} \frac{1}{s!} n^s \alpha^{(s)}(0) \tag{37}$$

Define the Fock space (up to an arbitrary function $f(n)$) by fixing the function $F(n)$ in the form $F(n) = n + 1$.

Then the operator of integer (22) defines the basis of the space $|n\rangle$, which are its eigenvectors:

$$N|n\rangle = n|n\rangle \tag{38}$$

Obviously,

$$N^s|n\rangle = n^s|n\rangle \tag{39}$$

where $s = 1, 2, ...$; $N^s = \underbrace{N \ldots N}_{s}$.

Thus, the basis $|n\rangle$ is also eigenvectors (eigenfunctions) of the operators $N^s$, while the parameters $n^s$ are their corresponding eigenvalues.

Taking into account (37) and (39), any operator (31) can be represented in this space as

$$\hat{A} = \sum_{s=0}^{\infty} \frac{1}{s!} \alpha^{(s)}(0) N^s \tag{40}$$

which is easily verified by substituting (40) into (31) by appealing to (39) and (37).

Relation (40) can be considered as a formal decomposition into a power series of an operator function

$$\hat{A} = \alpha(N) = \alpha(a^\dagger a) \tag{41}$$

which is the representation of the operator $\hat{A}$ through the operator of integer (22).

We now consider the general case of the representation of an operator $\hat{W}$ (30) satisfying condition (29) in terms of rising and lowering operators.

Obviously, we can write (30) in the form

$$\widehat{W}|n\rangle = \sum_k w_{nk}|k\rangle = \sum_{k<n} w_{nk}|k\rangle + \sum_{k>n} w_{nk}|k\rangle + w_{nn}|n\rangle \tag{42}$$

Taking into account (5) and (6), we can obtain

$$|k\rangle = \begin{cases} \dfrac{a^{\dagger^{k-n}}|n\rangle}{\prod_{s=n}^{k-1} f(s)}; k > n \\ \dfrac{a^{n-k}|n\rangle}{\prod_{s=k+1}^{n} \varphi(s)}; k < n \end{cases} \tag{43}$$

Then, taking into account (42) and (43), one obtains

$$\widehat{W}|n\rangle = \left(\sum_{k<n} w_{nk} \frac{a^{n-k}}{\prod_{s=k+1}^{n} \varphi(s)} + \sum_{k>n} \frac{a^{\dagger^{k-n}}}{\prod_{s=n}^{k-1} f(s)} + w_{nn}\right)|n\rangle$$

Thus, we can write

$$\widehat{W}(a, a^\dagger, n) = \sum_{k<n} w_{nk} \frac{a^{n-k}}{\prod_{s=k+1}^{n} \varphi(s)} + \sum_{k>n} \frac{a^{\dagger^{k-n}}}{\prod_{s=n}^{k-1} f(s)} + w_{nn} \tag{44}$$

The right hand side formally is a function of arguments $a$, $a^\dagger$ and $n$.

Taking into account (22), (38) and (41), we obtain the formal expression for operator $\widehat{W}$:

$$\widehat{W} = \widehat{W}(a, a^\dagger, a^\dagger a) \tag{45}$$

In this case, the operator of integer acts first and only then the operators $a$ and $a^\dagger$ act.

The choice of the function $f(n)$ in some cases can significantly simplify the form of the $\widehat{W}$ operator.

As an example we consider a simple birth-coagulation process [5] for which

$$w_{n,n+1} = \alpha n; \; w_{nn} = -\alpha n - \beta n(n-1); \; w_{n,n-1} = \beta n(n-1) \tag{46}$$

where $\alpha$ and $\beta$ are the constants. The rest $w_{nk} = 0$.

Then, taking into account (20), we can write (44) in the form

$$\widehat{W}(a, a^\dagger, n) = \beta a \frac{n(n-1)}{\varphi(n)} + \alpha a^\dagger \frac{n}{f(n)} - \alpha n - \beta n(n-1) \tag{47}$$

In the Doi-Peliti formalism (23) and (24), one obtains

$$\widehat{W}(a, a^\dagger, n) = \beta(a-n)(n-1) + \alpha(a^\dagger - 1)n \tag{48}$$

Taking into account (45), and substituting the operator of integer (22) instead of $n$ into (48) we obtain the expression for operator $\widehat{W}$:

$$\widehat{W} = \beta(a - a^\dagger a)(a^\dagger a - 1) + \alpha(a^\dagger - 1)a^\dagger a \tag{49}$$

After simple algebra, using (7), (8) and (16), one obtains the operator $\widehat{W}$ in the form [5]

$$\widehat{W} = \beta(1 - a^\dagger)a^\dagger a^2 + \alpha(a^\dagger - 1)a^\dagger a \tag{50}$$

**4. Vector representation of random observables**

We consider a *random observational process* $a$ which has outcomes $\ldots, a_0, a_1, a_2, \ldots$ being realized with probabilities $\ldots, p_0, p_1, p_2, \ldots$ which satisfy the normalization condition

$$\sum_n p_n = 1 \tag{51}$$

Then the mean value of this random observable is given by

$$\langle a \rangle = \sum_n a_n p_n \qquad (52)$$

Random observational process $a$ can be realized in different contexts. In our contextual probabilistic model a *context* for $a$-process are represented by a pair of vectors in the Fock space and its dual space:

$$|u\rangle = \sum_n c_n |n\rangle \qquad (53)$$

$$\langle u| = \sum_n b_n \langle n| \qquad (54)$$

where $b_n$ and $c_n$ are (in the general case) the complex numbers: the "coordinates" of a context[2]. We set the normalization condition, see (4), for the vectors $\langle u|, |u\rangle$, see (53), (54), representing a context,

$$\langle u|u\rangle = \sum_n c_n b_n = 1 \qquad (55)$$

We assume that condition (55) is another form of the normalization condition (51), its vector representation, by the pair of dual vectors. It means that

$$c_n b_n = p_n \qquad (56)$$

As shown in the previous sections, using the rising and lowering operators, one can always construct a linear operator $\hat{A}$ acting on vectors $|n\rangle$, such that the parameters $a_n$, which are realized in a random observation, are its eigenvalues:

$$\hat{A}|n\rangle = a_n |n\rangle \qquad (57)$$

In particular, such an operator $\hat{A}$ can always be expressed in terms of the operator of integer (22). Taking into account (56), (1) and (52), the mean value of the parameter $a$ in the random process under consideration can be represented as $\langle a \rangle = \langle u|\hat{A}|u\rangle$. From (57) it follows that $\hat{A}^s|n\rangle = a_n^s|n\rangle$, where $s = 1,2,...$; $\hat{A}^s = \underbrace{\hat{A}...\hat{A}}_{s}$. By definition,

$$\langle a^s \rangle = \sum_n a_n^s p_n \qquad (58)$$

Taking into account (1), (3), (4), (6) and (58), one can write

$$\langle a^s \rangle = \langle u|\hat{A}^s|u\rangle \qquad (59)$$

The mean value of a certain function $z(a)$, which depends on the parameter $a$, in the considered random process is determined by the relation

$$\langle z(a) \rangle = \sum_n z(a_n) p_n \qquad (60)$$

---

[2] In our terminology, "random observational process" is referred to the complete process of observation including generation of states of systems under observation. By using the operational terminology of quantum mechanics observational process $a$ includes both the preparation and measurement. One can say that the pair of dual vectors represents the preparation context. However, one has to be very careful with such terminology, because spitting of an observation in the preparation and measurement part is a delicate issue.

Using a power series $z(a) = \sum_{s=0}^{\infty} \frac{1}{s!} a^s z^{(s)}(0)$, from (60) taking into account (58) and (59) onr obtains $\langle z(a) \rangle = \sum_n \sum_{s=0}^{\infty} \frac{1}{s!} a_n^s z^{(s)}(0) p_n = \sum_{s=0}^{\infty} \frac{1}{s!} z^{(s)}(0) \sum_n a_n^s p_n = \sum_{s=0}^{\infty} \frac{1}{s!} z^{(s)}(0) \langle a^s \rangle = \sum_{s=0}^{\infty} \frac{1}{s!} \langle u | \hat{A}^s | u \rangle z^{(s)}(0)$. This expression can be rewritten in the form $\langle z(a) \rangle = \langle u | \sum_{s=0}^{\infty} \frac{1}{s!} z^{(s)}(0) \hat{A}^s | u \rangle$. As usual, we can introduce the formal operator $z(\hat{A}) = \sum_{s=0}^{\infty} \frac{1}{s!} z^{(s)}(0) \hat{A}^s$. Then, one obtains

$$\langle z(a) \rangle = \langle u | z(\hat{A}) | u \rangle \tag{61}$$

Thus, any statistical characteristic of the random process $u$ can be calculated (at least formally) using the dual vectors (53), (54) and the operator $\hat{A}$ (57).

Note that condition (56) only links the products $c_n b_n$ with the probabilities $p_n$, but in no way limits the parameters $c_n$ and $b_n$ themselves. This gives the freedom to choose the parameters $c_n$ and $b_n$, the product of which must satisfy condition (56). Different choice of these parameters leads to different Hilbert space representations by vectors $|u\rangle$ of the probability space (51). The special well known cases are:

- *the quantum probability space (second quantization of the bosonic field; $n = 0,1,2,...$):*

$$b_n = c_n^*, p_n = |c_n|^2 \tag{62}$$

or, taking into account (56)

$$c_n = \sqrt{p_n} \exp(i\beta_n), b_n = \sqrt{p_n} \exp(-i\beta_n) \tag{63}$$

- *the Doi-Peliti probability space [1-4] ($n = 0,1,2,...$):*

$$c_n = p_n \exp(i\beta_n), b_n = \exp(-i\beta_n) \tag{64}$$

where $\beta_n$ are the arbitrary parameters.

Summarizing (63) and (64), one can write

$$c_n = p_n^r \exp(i\beta_n), b_n = p_n^{1-r} \exp(-i\beta_n) \tag{65}$$

where $r$ is the constant.

In the general case, a probability space can be constructed by setting

$$c_n = p_n \Phi_n(P), b_n = \frac{1}{\Phi_n(P)} \tag{66}$$

where $\Phi_n(P)$ are the arbitrary (including complex-valued) functions depending on $P = (p_0, p_1, p_2, ...)$. Obviously, expressions (65) are the special case of expressions (66).

In the series of papers [7-12] (motivated by Feynman's work [47]), the phases in the state representation (63) were interpreted as representing contextuality, through the interference term modifying the formula of the total probability. The role of these phases as contextuality parameters is visible only through state superposition. (We remark that only relative phases contribute to the interference term.) In the Doi-Peliti probability space, the phases in the state representation (64), see also (65), can also be considered as parameters encoding contextuality; generally contextuality is represented by functions $\Phi_n(P)$, see section 5.

As an example, we consider the Doi-Peliti formalism as applied to the Kolmogorov's kinetic equation

$$\frac{dp_n}{dt} = \sum_k \lambda_{nk} p_k - p_n \sum_k \lambda_{kn} \tag{67}$$

We assume here that the parameters $\lambda_{nk}$ do not depend on the probabilities $\{p_n\}$, and also, for convenience without loss of generality, we accept $\lambda_{nn} = 0$.

We can rewrite equation (67) in the form

$$\frac{dp_n}{dt} = \sum_k w_{kn} p_k \tag{68}$$

Where $w_{kn} = \lambda_{nk}$ for $n \neq k$ and $w_{kn} = -\sum_s \lambda_{sn}$ for $n = k$.

We consider a vector in the Fock space in the Doi-Peliti formalism:

$$|u\rangle = \sum_n p_n |n\rangle \tag{69}$$

Differentiating (69) with respect to time, one obtains $\frac{d|u\rangle}{dt} = \sum_n \frac{dp_n}{dt} |n\rangle$.

Taking into account (68), one obtains

$$\frac{d|u\rangle}{dt} = \sum_n \sum_k w_{kn} p_k |n\rangle \tag{70}$$

Using the linear operator $\widehat{W}$, that satisfies condition (29), one can write equation (70) in the form

$$\frac{d|u\rangle}{dt} = \widehat{W} |u\rangle \tag{71}$$

The operator $\widehat{W}$ is expressed in terms of the rising and lowering operators by the relation (44) and (45).

## 5. Fock space and contextuality

In the past two decades, attempts have been made [6] *to describe a contextuality in random observation processes by interference of probabilities*, as is done in quantum theory. For this purpose, in fact, the representation of random observable as a normalized vector, see (63) in Fock space (63) and the Born rule were used. Contextuality (for non-compound systems) was expressed in the form of interference of probabilities as, for example, in the two-slit experiment. In mathematical terms such probabilistic interference was expressed as the violation of the classical formula of total probability [7-12]. Taking into account that (as shown above) the Fock space can be constructed in different ways, we conclude that *such a linear representation of random observation processes which depend on contextuality is also not unique.*

As for example, now consider the superposition of states in the Doi-Peliti probability space (64). Suppose there are two non-correlated processes $|u_1\rangle$ and $|u_2\rangle$ which are realized with probabilities $p_1$ and $p_2$, where $p_1 + p_2 = 1$. We assume that the phases of probabilities for these processes are equal $\beta_n^{(1)} = \beta_n^{(2)} = \beta_n$. We can introduce the vectors

$$|u\rangle = p_1|u_1\rangle + p_2|u_2\rangle = \sum_n \left(p_1 p_n^{(1)} + p_2 p_n^{(2)}\right) \exp(i\beta_n) |n\rangle$$

and

$$\langle u| = p_1\langle u_1| + p_2\langle u_2| = \sum_n (p_1 + p_2) \exp(-i\beta_n) \langle n| = \sum_n \exp(-i\beta_n) \langle n|$$

which describe the whole process. Note that in this case we can formally redefine the vectors $|n\rangle$: $|n\rangle' = \exp(i\beta_n) |n\rangle$. This means that we can consider $\beta_n \equiv 0$.

It is easy to verify that

$$\langle u|u\rangle = 1 \tag{72}$$

$$\langle u|\hat{A}|u\rangle = \langle a_1 \rangle p_1 + \langle a_2 \rangle p_2 = \langle a \rangle \tag{73}$$

Indeed,

$$\langle u|u\rangle = \sum_n \exp(-i\beta_n) \langle n| \sum_k \left(p_1 p_k^{(1)} + p_2 p_k^{(2)}\right) \exp(i\beta_n) |k\rangle = \sum_n \left(p_1 p_n^{(1)} + p_2 p_n^{(2)}\right) =$$
$$p_1 \sum_n p_n^{(1)} + p_2 \sum_n p_n^{(2)} = p_1 + p_2 = 1$$
$$\langle u|\hat{A}|u\rangle = \sum_n \exp(-i\beta_n) \langle n| \sum_k a_n \left(p_1 p_k^{(1)} + p_2 p_k^{(2)}\right) \exp(i\beta_n) |k\rangle = \sum_n a_n \left(p_1 p_n^{(1)} +$$
$$p_2 p_n^{(2)}\right) = p_1 \sum_n a_n p_n^{(1)} + p_2 \sum_n a_n p_n^{(2)} = \langle a_1 \rangle p_1 + \langle a_2 \rangle p_2$$

Thus, we see that in the Doi-Peliti probability space, see rule (64), for processes such that the phase parameters $\beta_n$ are equal; there is no interference of probabilities. This means that such a representation of the random processes does not take into account contextuality.

Consider now the general case: two observation-contexts $|u_1\rangle$ and $|u_2\rangle$ (and the corresponding dual vectors) which are realized with probabilities $p_1$ and $p_2$, where $p_1 + p_2 = 1$. These processes are described by the vectors (53), (54) constrained by (56). We will call the probabilities $p_n^{(1)}$ and $p_n^{(2)}$ the a priori probabilities of associated with these observation-contexts, i.e. probabilities of the realization of various events labeled by $n$ for these contexts, when contextual observations are performed independently of each other. We define the vector corresponding to the context of the combined context, as

$$|u\rangle = g_1|u_1\rangle + g_2|u_2\rangle = \sum_n \left(g_1 c_n^{(1)} + g_2 c_n^{(2)}\right)|n\rangle \tag{74}$$

and

$$\langle u| = q_1\langle u_1| + q_2\langle u_2| = \langle u| = \sum_n \left(q_1 b_n^{(1)} + q_2 b_n^{(2)}\right)\langle n| \tag{75}$$

where $g_1, g_2, q_1$ и $q_2$ are the parameters that satisfy the conditions

$$g_1 q_1 = p_1, \quad g_2 q_2 = p_2 \tag{76}$$

We introduce the probability $P_n$ of an event labeled by $n$ *for the combined context*. Obviously, the normalization condition must be satisfied

$$\sum_n P_n = 1 \tag{77}$$

Comparing (74) and (75) with (53) and (54), and taking into account (56), we come to the conclusion that

$$P_n = \left(g_1 c_n^{(1)} + g_2 c_n^{(2)}\right)\left(q_1 b_n^{(1)} + q_2 b_n^{(2)}\right) \tag{78}$$

Taking (56) and (76) into account, the expression (77) can be rewritten in the form

$$P_n = p_1 p_n^{(1)} + p_2 p_n^{(2)} + g_2 c_n^{(2)} q_1 b_n^{(1)} + g_1 c_n^{(1)} q_2 b_n^{(2)} \tag{79}$$

Probability (79) contains the interference term

$$R_n = g_2 c_n^{(2)} q_1 b_n^{(1)} + g_1 c_n^{(1)} q_2 b_n^{(2)} \tag{80}$$

which can describe the contextuality of the processes under consideration Taking (77) into account, the interference term must satisfy the condition

$$\sum_n R_n = 0 \tag{81}$$

By choosing the parameters $c_n^{(1)}, c_n^{(2)}, b_n^{(1)}, b_n^{(2)}, g_1, g_2, q_1$ and $q_2$, one can construct different contextuality models using linear Fock space. A special case is the quantum probability space (63), which are intensively discussed in connection with the contextuality problem [7-12]. At the same time, we see that other Fock spaces (e.g. Doi-Peliti probability space) can be considered as an acceptable tool when describing contextuality. These *models differ by the forms of the interference term*.

Consider the generalized Fock space (66) and define

$$g_{1,2} = \mu_{1,2} p_{1,2}; \quad q_{1,2} = \frac{1}{\mu_{1,2}} \tag{82}$$

where $\mu_{1,2}$ are the arbitrary (including complex-valued) functions.

Then the equation (79) for the "interference of probabilities" takes the form

$$P_n = p_1 p_n^{(1)} + p_2 p_n^{(2)} + (\mu_1/\mu_2)(\Phi_n^{(1)}/\Phi_n^{(2)}) p_1 p_n^{(1)} + (\mu_2/\mu_1)(\Phi_n^{(2)}/\Phi_n^{(1)}) p_2 p_n^{(2)} \tag{83}$$

We can introduce

$$\frac{\mu_1}{\mu_2} = \mu \exp(i\delta), \quad \frac{\Phi_n^{(1)}}{\Phi_n^{(2)}} = \Gamma_n \exp(i\gamma_n)$$

where $\mu, \delta, \Gamma_n$ and $\gamma_n$ are some real-valued parameters. In this case the expression (83) takes the form

$$P_n = p_1 p_n^{(1)} + p_2 p_n^{(2)} + \mu \Gamma_n p_1 p_n^{(1)} \exp(i(\gamma_n + \delta)) + \frac{1}{\mu \Gamma_n} p_2 p_n^{(2)} \exp(-i(\gamma_n + \delta)) \tag{84}$$

Taking into account that probabilities must be the real-valued parameters, we obtain the conditions

$$P_n = p_1 p_n^{(1)} + p_2 p_n^{(2)} + \left(\mu\Gamma_n p_1 p_n^{(1)} + \frac{1}{\mu\Gamma_n} p_2 p_n^{(2)}\right) \cos(\gamma_n + \delta) \tag{85}$$

$$\left(\mu\Gamma_n p_1 p_n^{(1)} - \frac{1}{\mu\Gamma_n} p_2 p_n^{(2)}\right) \sin(\gamma_n + \delta) = 0 \tag{86}$$

The equation (85) has two solutions:

$$\gamma_n + \delta \neq \pi k; \quad \mu\Gamma_n = \sqrt{\frac{p_2 \, p_n^{(2)}}{p_1 \, p_n^{(1)}}} \tag{87}$$

and

$$\gamma_n + \delta = \pi k \tag{88}$$

for any parameters $\mu\Gamma_n$, where $k = 0,1,2,\dots$.

In the case of (87), the probabilities (85) take the form

$$P_n = p_1 p_n^{(1)} + p_2 p_n^{(2)} + 2\sqrt{p_1 p_2 p_n^{(1)} p_n^{(2)}} \cos(\gamma_n + \delta) \tag{89}$$

Thus, in this case we have the quantum probability space (63) and quantum rule (the Born rule) for the interference of probabilities [7-12].

In the case of (88), we can consider only two possible values of the phase difference: $\gamma_n + \delta = 0$ and $\gamma_n + \delta = \pi$ which correspond to $\cos(\gamma_n + \delta) = \pm 1$. In the first case, random processes are in the same phase, while in the second case they are in antiphase.

For some $n$, the phase difference can take the value $\gamma_n + \delta = 0$, while for the others, the value $\gamma_n + \delta = \pi$, however, in general, the normalization condition (77) should be satisfied.

Thus, the case $\gamma_n + \delta = 0$ corresponds to an increase in the probability $P_n$:

$$P_n = (1 + \mu\Gamma_n) p_1 p_n^{(1)} + \left(1 + \frac{1}{\mu\Gamma_n}\right) p_2 p_n^{(2)} \tag{90}$$

while the case $\gamma_n + \delta = \pi$ corresponds to an decrease in the probability $P_n$:

$$P_n = (1 - \mu\Gamma_n) p_1 p_n^{(1)} + \left(1 - \frac{1}{\mu\Gamma_n}\right) p_2 p_n^{(2)} \tag{91}$$

The case $\mu\Gamma_n = 1$ corresponds to the Doi-Peliti formaliusm (64).

Note that in the case of (88), a priori probabilities $p_1, p_2, p_n^{(1)}$ and $p_n^{(2)}$ enter the interference term linearly. The formulas (90) and (91) of total probability with such interference terms are closer to the classical formula of total probability than the formula with the quantum interference term that occurs in the quantum probability space (63), when not the probabilities, but the probability amplitudes, i.e. the "wave functions", are summarized (see (89)).

## 6. Concluding remarks

We construct the calculi of creation-annihilation operators acting in generalized Fock spaces. Such spaces are used for the linear space representations of contextual probabilities. As the special cases, we reproduce quantum probability based on complex Hilbert space and Doi-Peliti probability space [1-4]. Such representations give place to generalizations of the formula of total probability, see (89), (90), and (91). The interference terms can be used to estimate the degree of contextuality. We remark that already Feynman [47] connected quantum interference of probabilities in the two-slit experiment with its contextual structure. As was shown in [7-12], generally contextual structure of experiments leads to perturbations of the classical formula of total probability. In this paper, we generate concrete forms of such perturbations on the basis of the representations in generalized Fock spaces.

The Fock space representations and calculi of creation-annihilation operators in these spaces are used to construct the linear space representations of the classical kinetic equations in the spirit of Doi-Peliti formalism. As was emphasized by Peliti [4]: "Such a formulation has the advantage of easily allowing for an estimation of the validity of approximations, or for building up a systematic perturbation scheme, which may eventually lead to the application of renormalization group methods… The scope of these methods widely extends beyond the realm of aggregation processes… they help in understanding the relation between classical and quantum statistics in laser-like systems, as well as how they provide new points of view on the nature of quasi-deterministic approximations in chemical systems."

Note, that there is a connection of approach under consideration with the physics of nonlinear vibrations and mathematical aspects of quantum groups. In particular the Fock state formulas considered in this paper are related with formalism of f-oscillators [48] and corresponding quantum algebra formalism.

Quantum objects have the property of indistinguishability, which determines the statistics of their behavior. In this connection, the question about the inclusion of the generalized versions of quantum indistinguishability into formalism under consideration arises. This issue is partially discussed in [43] by one of the co-authors of this paper. As shown in [43], there are objects outside of physics that possess the 'bosonic' and 'fermionic' properties and therefore must obey the corresponding statistics.

In future publications we plan to proceed to concrete applications of our formalism to physics, biology, and decision making.

**Acknowledgments**

This work was done on the theme of the State Task No. AAAA-A17-117021310385-6. Funding was provided in part by the Tomsk State University competitiveness improvement program. This work was financially supported by Government of Russian Federation, Grant 08-08 and by Ministry of Education and Science of the Russian Federation within the Federal Program "Research and development in priority areas for the development of the scientific and technological complex of Russia for 2014-2020", Activity 1.1, Agreement on Grant No. 14.572.21.0008 of 23 October, 2017, unique identifier: RFMEFI57217X0008.